\documentclass[submission,copyright,creativecommons]{eptcs}
\providecommand{\event}{CompMod 2011}
\usepackage{graphicx}
\usepackage{subfigure}
\usepackage{amssymb}
\usepackage{color}
\usepackage{url}
\usepackage{multirow}

\newtheorem{definition}{\bf Definition}[section]

\title{A Study of the PDGF Signaling Pathway with PRISM}
\author{
Qixia Yuan
\institute{
Computer Science and Communications \\
University of Luxembourg, Luxembourg}
\institute{
School of Computer Science and Technology\\
Shandong University, China}
\and
Jun Pang
\institute{
Computer Science and Communications\\
University of Luxembourg, Luxembourg}
\and
Sjouke Mauw
\institute{
Computer Science and Communications\\
University of Luxembourg, Luxembourg}
\and
Panuwat Trairatphisan
\institute{
Life Sciences Research Unit\\
University of Luxembourg, Luxembourg}
\and
Monique Wiesinger
\institute{
Life Sciences Research Unit\\
University of Luxembourg, Luxembourg}
\and
Thomas Sauter
\institute{
Life Sciences Research Unit\\
University of Luxembourg, Luxembourg}
}

\def\titlerunning{A Study of the PDGF Signaling Pathway with PRISM}
\def\authorrunning{Yuan {\em et al.}}
\begin{document}
\maketitle

\begin{abstract}
In this paper, we apply the probabilistic model checker PRISM
to the analysis of a biological system -- the Platelet-Derived Growth Factor (PDGF) signaling pathway,
demonstrating in detail how this pathway can be analyzed in PRISM.
We show that quantitative verification
can yield a better understanding of the PDGF signaling pathway.
\end{abstract}

\section{Introduction}
\label{sec:introduction}

Biological systems consist of separated components,
which interact to influence each other and therefore the whole system's behavior.
The field of systems biology aims to understand such complex interactions 
from a systematic view.
Due to the similarity between biological systems and
complex distributed/reactive systems studied in computer science~\cite{RS02},
modeling and analyzing techniques developed in the field of formal methods
can be applied to biological systems as well~\cite{BFF+09}.
Due to efficient verification techniques,
formal methods can analyze large systems exhibiting complex behaviors
-- this process is typically supported by automatic computer tools.
Clearly, this gives formal methods an advantage,
as \emph{in silico} experiments are much easier to perform than \emph{in vitro} experiments
for the aim of analyzing and understanding biological systems.
During the last decade, there has been a rapid and successful development
in applying formal methods to systems biology --
new formalisms are developed for systems biology
to create models for biological phenomena,
new algorithms and tools are specially designed and tailored for the analysis of such models
(e.g., see~\cite{FH07,SFB+08,FP10}).

In this paper, we explore the usage of model checking
for biological systems.
Model checking is referred to as the automatic process of checking whether
a system model satisfies a given specification (expressed as a temporal logic formula),
by exhaustively exploring all possible executions of the system.
This differs from simulation-based techniques,
which only study a subset of the executions.
More specifically, we focus on the  probabilistic model checking approach,
first introduced by Hart, Sharir and Pnueli~\cite{HSP83},
as biological systems usually have complicated stochastic behaviors.
This technique is well-established and widely used for ascertaining the correctness of real-life systems,
including distributed systems and communication protocols.
In probabilistic model checking, systems are normally represented by Markov chains
or Markov decision processes.
Properties of the models are expressed in quantitative extensions of temporal logics.
Probabilistic verification has gained notable success in
analyzing probabilistic systems including biological signaling pathways (e.g., see~\cite{KNP09b,KNP10a}).

We use the probabilistic model checker PRISM~\cite{KNP10a} to
yield a better understanding of the PDGF (Platelet-Derived Growth Factor) signaling pathway.
PDGF, described approximately 30 years ago as a major mitogenic component of whole blood~\cite{YUK03},
is a growth factor that regulates cell growth and division.
It promotes angiogenesis and also preserves vascular integrity through the recruitment of pericytes to endothelial tubes.
Clinical studies reveal that aberrant expression of PDGF and its receptor is often associated with a variety of disorders
such as atherosclerosis, fibroproliferative diseases and most importantly, neoplasia~\cite{YUK03}.
Deregulation of the PDGF signaling pathway plays a critical role in the development of many types of human diseases such as
gastrointestinal stromal tumor and hypereosinophilic syndrome~\cite{GBLG94,KTM+96,YUK03,JDS+03,ML06}.
Based on intensive literature review,
we have built the PDGF signal transduction model in ODE (Ordinary Differential Equation) format.
The essential part of the PDGF signaling pathway contains the coupling of PDGF ligand to its receptor {\tt PDGFR},
the negative regulatory mechanism on {\tt PDGFR}
and the activation of two main downstream signaling pathways, i.e.,  MAPK (Mitogen-Activated Protein Kinase) and PI3K/Akt pathways.
In addition, there also exist positive and negative crosstalk interactions
between different downstream signaling pathways
(more details on the PDGF signaling pathway can be found in Section~\ref{sec:pdgf}).
In our study, there are three main goals:
(1) analyze the dynamics of PDGF induced signaling,
(2) analyze the influence of the crosstalk reactions
and (3) analyze the importance of individual reaction on downstream signaling molecules.
The first two can be used to check whether the constructed signaling pathway
is consistent with respect to biological data,
while the last one can lead us to some prediction.
We have achieved these goals by quantitative verification using PRISM.

\bigskip\noindent
\emph{Related work.}
PRISM has been used for analyzing a variety of biological systems.
In~\cite{HKN+08}, PRISM is used to analyze the FGF (Fibroblast Growth Factor) signaling pathway.
Although only a model corresponding to a single instance of the pathway is built,
it is still rich enough to explain the roles of the components in the pathway and how they interact.
In~\cite{KNP08}, PRISM is used to study the MAPK cascade.
The authors explain how the biological pathway can be modeled in PRISM
and how this enables the analysis of a rich set of quantitative properties.

Jha {\em et al.}~\cite{KCLLPZ09} present the first algorithm for performing statistical model
checking using Bayesian sequential hypothesis testing
and test the performance of the algorithm on the FGF signaling pathway and several others.
Schwarick and Heiner~\cite{SH09} give an interval decision diagram based approach to CSL model checking.
They aim for efficient verification of biochemical network models
and apply their approach to the MAPK cascade.
Dra\'{a}bik {\em et al.}~\cite{DMM10} apply the technique modular verification
to analyzing the {\it lac} operon regulation.

Apart from probabilistic (or stochastic) model checking,
other formal techniques have been used for studying biological systems in recent years as well.
For instance, the process algebra bio-PEPA is used to model and analyze the NF-kappaB pathway~\cite{CDHH10};
Petri nets are used for quantitative predictive modeling of {\em C.elegans} vulval development~\cite{BKF+09};
and the rule-based modeling language Kappa is used to model
epigenetic information maintenance~\cite{KDB09}.
More related work, especially case studies of applying formal methods in systems biology,
can be found in~\cite{SFB+08,FP10}.

\bigskip\noindent
\emph{Outline of the paper.}
In Section~\ref{sec:prism}, we give an overview of probabilistic model checking and the tool PRISM.
Section~\ref{sec:pdgf} describes the PDGF signaling pathway.
In Section~\ref{sec:modellingandproperty}, we build a model in PRISM for the PDGF signaling pathway
and describe several properties of the model that we are interested in.
Our verification results are given in Section~\ref{sec:results}.
Finally, we draw conclusion of this paper and discuss some future work in Section~\ref{sec:conclusion}.

\section{Probabilistic Model Checking and PRISM}
\label{sec:prism}

We briefly introduce probabilistic verification
and the probabilistic model checker -- PRISM~\cite{KNP10a}.

\subsection{CTMC and CSL}
Probabilistic model checking is a variant of model checking,
which aims at analyzing the correctness of finite state systems
with focus on quantitative aspects.
Model checking of a system requires two inputs:
a formal description of the system,
which is usually given in a high level modeling formalism (e.g., Petri nets or process algebra)
and a specification of the system properties,
which is usually given as temporal logic (e.g., CTL or LTL) formulas.
After accepting the two inputs,
a model checking tool then can verify
whether the system satisfies the desired properties
and give counter-examples if the system does not satisfy a certain property,
by exploring all possible behaviors of the system exhaustively.
As the word ``probabilistic" indicates,
probabilistic model checking focuses on systems with stochastic behaviors.
Instead of asking the model checker ``will the molecule become active in the end?",
we can ask ``what is the probability of the molecule being active at the steady state?"
or ``what is the probability of the molecule being active at time instant $t$?".
In probabilistic model checking,
systems are normally represented by Markov chains or Markov decision processes.
In this paper, we use continuous-time Markov chains (CTMCs),
to build the signaling pathway models.

A CTMC can model both (continuous) real time and probabilistic choice
by assigning rates at transitions between states.
The formal definition of a CTMC is given as follows.

\begin{definition}
Let $\mathbb{R}_{\geq0}$ denote the set of non-negative reals
and $AP$ be a fixed finite set of atomic propositions.
A CTMC is a tuple $(S, R, L)$ where:
\begin{itemize}
\item $S$ is a finite set of states;
\item $R : S\times S \rightarrow \mathbb{R}_{\geq0}$ is a transition rate matrix;
\item $L : S \rightarrow 2^{AP}$ is a labeling function which associates each state with a set of atomic propositions.
\end{itemize}
\end{definition}
The transition rate matrix $R$ assigns rates to each pair of states,
which are used as parameters of the exponential distribution.
A transition can occur between two states $s$ and $s'$ if $R(s, s')> 0$,
and the probability of the transition being triggered within
$t$ time-units equals to $1-e^{R(s,s')\cdot t}$.
If $R(s, s')> 0$ for more than one state $s'$,
the first transition to be triggered determines the next state.
Therefore, the choice of the successor state of $s$ is probabilistic.
The time spent in state $s$ before any such transition occurs is
exponentially distributed with $E(s)=\sum_{s'\in S}R(s,s')$.
Hence, the probability of moving to state $s'$ is $\frac{R(s,s')}{E(s)}$,
i.e., the probability that the delay of going from $s$ to $s'$
``finishes before" the delays of any other outgoing transition from $s$.
A path in a CTMC is a sequence $\sigma$ in the form of
$s_0t_0s_1t_1\cdots$ with $R(s_i,s_{i+1})>0$ and $t_i\in \mathbb{R}_{\geq0}$
for all $i\geq 0$.
The amount of time spent in $s_i$ is denoted by $t_i$.

Corresponding to CTMC models,
we use Continuous Stochastic Logic (CSL) to specify properties of built models.
CSL, originally introduced by Aziz {\em et al.}~\cite{ASSB00},
provides a powerful means to specify both path-based and
traditional state-based performance measures on CTMCs.
\begin{definition}
The syntax of CSL is given as follows:
\[ \phi\ ::=\ true \mid a \mid \neg \phi \mid \phi \land \phi \mid P_{\sim p}[\phi U^I \phi] \mid S_{\sim p}[\phi] \]
where $a$ is an atomic proposition,
$\sim \in \{<, \leq, \geq, > \}$, $p \in [0, 1]$,
$I$ is an interval of $\mathbb{R}_{\geqslant 0}$ and $r,t \in \mathbb{R}_{\geq 0}$.
\end{definition}
CSL formulas are evaluated over the states of a CTMC.
CSL includes the standard operators from propositional logic:
$true$ (satisfied in all states);
atomic propositions ($a$ is true in states which are labelled with $a$);
negation ($\neg \phi$ is true if $\phi$ is not);
and conjunction ($\phi_1 \land \phi_2$ is true if both $\phi_1$ and $\phi_2$ are true).
Other standard boolean operators
can be derived from these in the usual way.
CSL also includes two probabilistic operators, $P$ and $S$,
both of which include a probability bound $\sim p$.
A formula $P_{\sim p} [\psi]$ is true in a state $s$ if the probability
of the path formula $\psi$ being satisfied from state $s$ meets the bound $\sim p$.
In this paper, we use a single type of path formula,
$F^I \phi \equiv true\, U^I \phi$, called an \textit{eventual} formula,
which is true for a path $\sigma$ if $\phi$ eventually becomes true
for some time instant $t\in I$.
Particularly, if the time interval is set to zero,
e.g. $F^{[t,t]} \phi$, the formula is true for a path $\sigma$
if $\phi$ becomes true at time instant $t$.
The $S$ operator is used to specify steady-state behavior of a CTMC.
More precisely, $S_{\sim p} [\psi]$ asserts that
the steady-state probability of being in a state satisfying $\psi$ meets the bound $\sim p$.

\subsection{The model checker PRISM}
\label{ssec:PRISM}
PRISM~\cite{KNP10a} is a model checking tool developed at the universities of Birmingham and Oxford.
It allows one to model and analyze systems containing stochastic behaviors.
PRISM supports three kinds of models: discrete-time Markov chains (DTMCs),
continuous-time Markov chains (CTMCs) and Markov decision processes (MDPs).
Analysis is performed through model checking such systems
against properties written in the probabilistic temporal logics PCTL
if the model is a DTMC or an MDP, or CSL in the case of a CTMC,
as well as their extensions for quantitative specifications and costs/rewards.

In PRISM a model is composed of a number of modules
that contain variables and can interact with each other.
The values of the variables at any given time constitute the state of the module,
and the local states of all modules decide the global state of the whole model.
The behavior of a module, normally the changes in states which it can undergo,
is specified by a set of guarded commands of the form:
\[ [a]\ g \rightarrow r : u; \]
$a$ is an action label in the style of process algebra,
which introduces synchronization into the model.
It can only be performed simultaneously by all modules that have
an occurrence of action label $a$.
If a transition does not have to synchronize with other transitions,
then no action label needs to be provided for this transition.
The symbol $g$ is a predicate over all the variables in the system.
A guarded command means that if the guard $g$ is true,
the system is updated according to $u$
with rate $r$, which is corresponding to the transition rate of CTMC.
A transition updates the value of variables by giving their new {\em primed} value
with respect to their {\em unprimed} value.

PRISM models can be augmented with information about rewards (or equivalently, costs).
The tool can analyze properties which relate to the expected values of these rewards.
A CTMC in PRISM can be augmented with two types of rewards:
{\em state reward} associated with states
which are accumulated in proportion to the time spent in the state, and
{\em transition reward} associated with
transitions which are accumulated each time the transition is taken.
CSL is extended with quantitative costs/rewards as well,
which is quite useful in analyzing the quantitative properties of a biological system,
by introducing the $R$ operator:
\[R\ ::=\ R_{\sim r}[I^{=t}] \mid R_{\sim r}[C^{\leq t}] \mid R_{\sim r}[F \phi] \mid R_{\sim r}[S] \]
where $\sim \in \{<,\leq, \geq, > \}$, $r,t \in \mathbb{R}_{\geq 0}$ and $\phi$ is a CSL formula.
Intuitively, a state $s$ satisfies $R_{\sim r}[I^{=t}]$ if
from $s$ the expected state reward at time instance $t$ meets the bound $\sim r$;
a state $s$ satisfies $R_{\sim r}[C^{\leq t}]$ if
the expected reward accumulated up until $t$ time units past satisfies $\sim r$;
a state $s$ satisfies $R_{\sim r}[F \phi]$ if
from $s$ the expected reward accumulated before a state satisfying $\phi$ is reached meets the bound $\sim r$;
and a state $s$ satisfies $R_{\sim r}[S]$ if
from $s$ the long-run average expected reward satisfies $\sim r$.

It is often useful to take a quantitative approach to probabilistic model checking,
computing the actual probability that some behavior of a model is observed,
rather than just verifying whether or not the probability is above or below a given bound.
Hence, PRISM allows the $P$ and $S$ operators in CSL to take the following form:
$P_{=?}[\psi]$ and $S_{=?}[\psi]$.

\section{The PDGF Signaling Pathway}
\label{sec:pdgf}

Cell signaling is part of a complex system in cellular communication.
It allows the cells to activate a large number of signaling molecules and to regulate their activity.
In order to transfer a regulatory signal upon reception of a triggering stimulus,
the signal is transformed into a chemical messenger within the signaling cell,
e.g., via transfer of a phosphate group (phosphorylation)~\cite{Kra08}.
For further details on cell signaling see, for example,~\cite{Bha03,Kra08}.

Platelet-Derived Growth Factor (PDGF), described approximately 30 years ago
as a major mitogenic component of whole blood~\cite{YUK03}
is a growth factor that regulates cell growth and division.
By binding to its receptor ({\tt PDGFR}),
it regulates many biological processes such as migration, survival and proliferation~\cite{HWW85}.
{\tt PDGFR} is a receptor tyrosine kinase,
which in general transfer upstream signals to many downstream signaling pathways by phosphorylation.
Up to now, five PDGF ligands are known, PDGF-AA, -AB, -BB, -CC, -DD,
interacting with three different types of {\tt PDGFR} complexes,
{\tt PDGFR}-$\alpha\alpha$, -$\alpha\beta$ and -$\beta\beta$.
Each of the {\tt PDGFR} subtypes has a different affinity to the different PDGF ligands~\cite{TK04}.

\begin{figure}[!ht]
\begin{center}
\includegraphics[scale=0.6]{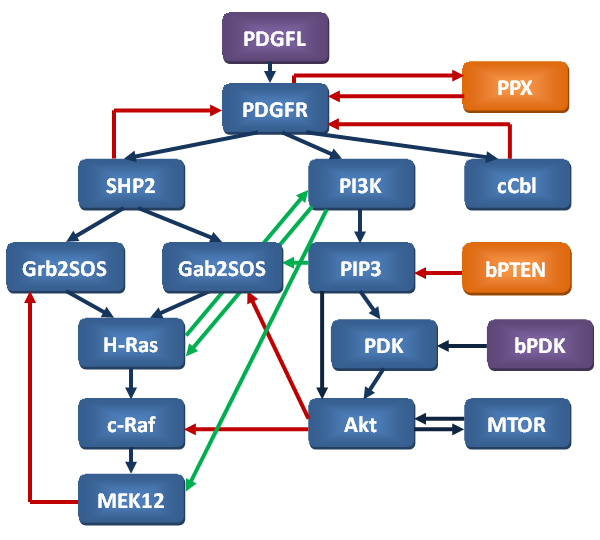}
\caption{The extracted PDGF signaling pathway (blue arrows: main pathway, green arrows: positive crosstalk, red arrows: negative regulatory)\label{fig:pdgf}}
\end{center}
\end{figure}

After {\tt PDGFRs} couple with their respective ligands,
phosphorylation of the receptor at specific tyrosine residues will occur,
thus enabling binding of signaling enzymes including {\tt Src}, phosphatidylinositol 3 kinase ({\tt PI3K}),
phospholipase C$\gamma$ (PLC$\gamma$) and {\tt SHP2} in the MAPK pathway at specific binding sites.
The recruitment of these signaling enzymes to {\tt PDGFR} is mediated via an intrinsic {\tt SH2} domain.
The translocation of {\tt PI3K} and PLC$\gamma$ to the plasma membrane also increases their accessibility to their respective substrates.
Moreover, recent findings suggest that {\tt PDGFR} also has potential binding sites for CrkL~\cite{YHM98},
which will activate {\tt Rap1} to positively influence {\tt c-Raf}  in the MAPK pathway~\cite{GAP10},
for Signal Transducer and Activator of Transcription (STAT),
which might regulate the signal in parallel to the JAK-STAT pathway~\cite{VPC98}
and also for {\tt cCbl}, which promotes ubiquitination of {\tt PDGFR}. {\tt cCbl} is also considered to
be one of the negative regulatory molecules in PDGF signal transduction~\cite{JWL99}.
Based on intensive literature review,
we have built a PDGF signal tranduction model in ODE format.
This full model comprises 35 molecules.
The essential parts of the PDGF signaling pathway which are the coupling of PDGF ligand to {\tt PDGFR},
the negative regulatory feedbacks on {\tt PDGFR}
and the activation of two main downstream signaling pathways, the MAPK and PI3K/Akt  pathways,
are extracted for analysis in PRISM (as shown in Figure~\ref{fig:pdgf}).
This simplified model contains only 17 molecules, and the analysis in this paper
focuses on this simplified model.

Figure~\ref{fig:pdgf} describes how signals are transduced in the PDGF pathway
by activating or deactivating specific downstream pathways signaling molecules respectively.
In this model, there are three inputs which are {\tt PDGFL} (PDGF ligand), {\tt bPTEN}, and {\tt bPDK}.
{\tt PDGFL} is the node which represents the upstream molecule activating the whole network.
{\tt PPX}, and {\tt bPTEN} are nodes which represent phosphatase enzymes in the cytoplasm
that negatively regulate their targets.
Lastly, {\tt bPDK}, standing for basal activity of {\tt PDK},
is the node that constantly gives a basal additional input to {\tt PDK}  node.
This node is always active in order to activate the survival pathway
to counteract apoptotic signal and keep the cell survive at a basal level.
There are three different types of arrows in the network: blue arrows, green arrows and red arrows.
The blue arrows represent the main activating interactions,
indicating the two main downstream signaling pathways in the network.
The MAPK pathway covers {\tt SHP2}, {\tt Grb2SOS}, {\tt GabSOS}, {\tt Ras}, {\tt c-Raf} and {\tt MEK12}.
These molecules play a major role in the cellular proliferative circuit~\cite{HW11}.
The PI3K/Akt  pathway covers the molecules {\tt PI3K}, {\tt PIP3}, {\tt PDK}, {\tt bPTEN}, {\tt bPDK} and {\tt Akt},
also being important in viability circuit~\cite{HW11}.
The green arrows represent positive crosstalk interactions to the molecules the arrows point to.
Lastly, the red arrows represent either negative crosstalk interactions or other negative regulatory interactions.
The molecules will become active after they have been activated by either blue or green arrows.
In contrary, the molecules will become inactive
after they have been deactivated by the red arrows or the basal phosphatase activities in the cell (not shown in the figure).

{\tt PDGFR} can be activated by {\tt PDGFL}.
The active {\tt PDGFR} in turn activates three downstream molecules
which are {\tt SHP2}, {\tt PI3K}  and {\tt cCbl}.
Both {\tt SHP2} and {\tt cCbl} assert a negative feedback to PDGF making it inactive.
The three blue arrows connecting {\tt PDGFR}  to these downstream signaling molecules,
so called \emph{mutant arrows},
are the targets of system interventions both experimentally and computationally.
The experimental intervention can be performed by introducing a point mutation from tyrosine to phenylalanine
at the specific recruitment site for the downstream signaling enzyme
(Y720F for {\tt SHP2}  recruitment site, YY731/742FF for {\tt PI3K}  recruitment site,
and Y1018F for {\tt cCbl}  recruitment site), leading to the loss of signal capacity of the respective signaling pathway~\cite{YGM+94,BGK96,RYD+07}.
Thus, the result of computational simulation such as
the relative activities at the steady state of downstream signaling molecules from these respective mutants
can be validated experimentally in biological laboratories.

\begin{figure}
\begin{center}
\includegraphics[width=\textwidth]{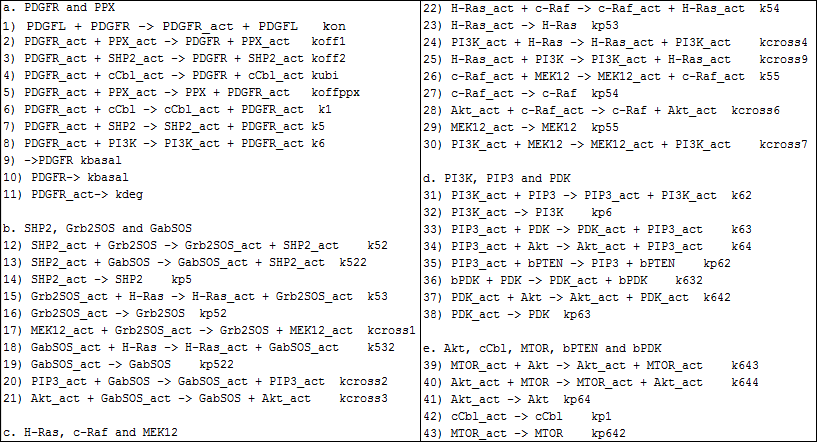}
\caption{List of biochemical reactions in the PDGF signaling pathway ODE model\label{fig:reactions} }
\end{center}
\end{figure}

Figure~\ref{fig:reactions} contains the list of model reactions.
Each node (molecule) is simplified to be in two states, either \emph{inactive} or \emph{active}
(indicated by the suffix {\tt \_act} in the figure).
All the reactions except for reactions 9, 10 and 11 describe the molecules changing between the two states;
while reactions 9, 10 and 11 indicate the basal production, the basal degradation
and the internalization following activation of {\tt PDGFR}.
For instance, reaction 2 describes that an active {\tt PPX} gives a negative feedback to {\tt PDGFR},
making it inactive.
The parameters used in the model (all starting with letter {\tt k}) in each reaction
are the relative reaction rate constants normalized to the maximum value in each set of reactions.
In this case, the values of parameters are ranging from 0 to 1.
The initial set of parameters' values used in this model
are derived from literature (e.g.,~\cite{GBLG94,KTM+96,YUK03,JDS+03,ML06}) and
our own experimental observations of {\tt PDGFR}$\alpha$ mutants.
Further parameter optimization based on experimental data is ongoing,
especially focusing on the unknown crosstalk reaction strength.
However, in this paper, the initial set of parameters has been used.
This might not fully correlate to the actual biological system.
Therefore, we focus in our analysis on the more qualitative aspects already conserved by the network structure.
%
%

\section{Modelling and Property Specifications in PRISM}
\label{sec:modellingandproperty}

\subsection{PRISM model}
\label{ssec:modelling}
We now describe how to build a PRISM model for the PDGF signaling pathway as presented in the previous section.
Our model represents a single instance of the signaling pathway,
meaning there can be at most one element of each molecule.
In the single instance model, the molecules' steady state,
which is expressed as a probability in PRISM,
corresponds to the molecule density in the real-life experiments.

\begin{figure}[!ht]
  \subfigure[PRISM module for {\tt PDGFR}\label{fig:modulePDGFR}]{
            \begin{minipage}[c]{0.99\textwidth}
            \centering\footnotesize
\fbox{\parbox{280pt}{

\vspace{2mm}
\hspace{5mm}{\tt module} {\it PDGFR}

\vspace{3mm}
\hspace{10mm}{\it PDGFR} : [0..2] init 0; //{\it 0 -- inactive}; {\it 1 -- active};  {\it 2 -- degraded}

\hspace{10mm}[] {\it PDGFL} \& {\it PDGFR}=0 $\rightarrow$ {\it kon} : {\it (PDGFR'}=1);

\hspace{10mm}[{\it bkoff1}] {\it PDGFR}=1 $\rightarrow$ {\it koff1} : {\it (PDGFR'}=0);

\hspace{10mm}[{\it bkoff2}] {\it PDGFR}=1 $\rightarrow$ {\it koff2} : {\it (PDGFR'}=0);
	
\hspace{10mm}[{\it bkubi}] {\it PDGFR}=1 $\rightarrow$ {\it kubi} : {\it (PDGFR'}=0); 	
	
\hspace{10mm}[{\it bkoffppx}] {\it PDGFR}=1 $\rightarrow$ {\it koffppx} : {\it (PDGFR'}=1);
	
\hspace{10mm}[{\it bk1}] {\it PDGFR}=1 $\rightarrow$ {\it k1} : {\it (PDGFR'}=1);
	
\hspace{10mm}[{\it bk5}] {\it PDGFR}=1 $\rightarrow$ {\it k5} : {\it (PDGFR'}=1);
	
\hspace{10mm}[{\it bk6}] {\it PDGFR}=1 $\rightarrow$ {\it k6} : {\it (PDGFR'}=1);
	
\hspace{10mm}[] {\it PDGFR}=0 $\rightarrow$ {\it kbasal} : {\it (PDGFR'}=2); //{\it PDGFR degraded}
	
\hspace{10mm}[] {\it PDGFR}=2 $\rightarrow$ {\it kbasal} : {\it (PDGFR'}=0);
	
\hspace{10mm}[] {\it PDGFR}=1 $\rightarrow$ {\it kdeg} : {\it (PDGFR'}=2); //{\it PDGFR$\_$act degraded}

\vspace{3mm}
\hspace{5mm}{\tt endmodule}

\vspace{2mm}
} }

\vspace{2mm}
            \end{minipage}}

\subfigure[PRISM model for {\tt PPX}\label{fig:modulePPX}]{
            \begin{minipage}[c]{0.48\textwidth}
            \centering\footnotesize
\fbox{\parbox{150pt}{

\vspace{2mm}
\hspace{5mm}{\tt module} {\it PPX}

\vspace{3mm}
\hspace{10mm}{\it PPX} : [0..1] init 1;

\hspace{10mm}[{\it bkoff1}] {\it PPX=1} $\rightarrow$ ({\it PPX'}=1);

\hspace{10mm}[{\it bkoffppx}] {\it PPX=1} $\rightarrow$ ({\it PPX'}=0);

\vspace{3mm}
\hspace{5mm}{\tt endmodule}

\vspace{2mm}
} }

\vspace{2mm}
            \end{minipage}}
\subfigure[PRISM rewards\label{fig:rewards}]{
            \begin{minipage}[c]{0.48\textwidth}
            \centering\footnotesize
\fbox{\parbox{120pt}{

\vspace{2mm}
\hspace{5mm}{\tt rewards} $``PDGFRactive"$

\vspace{3mm}
\hspace{10mm}{\it PDGFR}=1:1;

\vspace{3mm}
\hspace{5mm}endrewards

\vspace{2mm}
} }

\vspace{2mm}
            \end{minipage}}
\caption{PRISM modules and rewards\label{fig:model}}
\end{figure}

Each of the nodes (or molecules) of the pathway except for the {\tt PDGFL} in Figure~\ref{fig:pdgf}
is represented by a separate PRISM module.
Since {\tt PDGFL}, {\tt bPTEN} and {\tt bPDK} remain the same after the reactions they are involved,
we set them as const boolean values {\tt true}.
Figures~\ref{fig:modulePDGFR} and ~\ref{fig:modulePPX}  show the modules for {\tt PDGFR} and {\tt PPX}.
In each of the modules, the status of the molecule
is represented by a variable with the same name as the module.
The variables can have values of either 0 or 1
({\tt PDGFR} is an exception, since it can have value 2 since it can degrade),
corresponding to the two states,  inactive and active, of a molecule.
Each command in the PRISM module represents a reaction in Figure~\ref{fig:reactions}.
Interactions of multiple molecules are implemented by the synchronization between modules.
More precisely, the same label is given to the commands
which require synchronization in PRISM modules.
For example, in Figures~\ref{fig:modulePDGFR} and ~\ref{fig:modulePPX},
there are commands with label {\tt bkoff1} in both of the modules {\tt PDGFR} and {\tt PPX}.
The two commands are used to model the reaction (2) in Figure~\ref{fig:reactions},
which involves both {\tt PDGFR} and {\tt PPX}.
It guarantees that the two commands (corresponding to one reaction) can only occur
when both guards are satisfied.
The reaction rate is assigned by the command in module {\tt PDGFR}
and hence the reaction rate of the command in module {\tt PPX}  is omitted.
Totally we have modeled all the 17 molecules in 14 PRISM modules
({\tt PDGFL}, {\tt bPTEN} and {\tt bPDK} are modeled as a constant).

As mentioned in Section~\ref{ssec:PRISM},
PRISM models can be augmented with information about rewards.
We construct rewards to calculate the time for a molecule being active.
Figure~\ref{fig:rewards} shows the rewards for calculating the active state of {\tt PDGFR}.
Each time {\tt PDGFR} is in active state,
one is added to the total time of {\tt PDGFR} being active.
Similarly, we build rewards structures for other molecules as well,
including {\tt SHP2}, {\tt Ras}, {\tt MEK12}, {\tt PIP3} and {\tt Akt}.

\subsection{Property specifications}
\label{ssec:properties}

There are three main goals for this study:
(1) analyze the dynamics of PDGF induced signaling,
(2) analyze the influence of the crosstalk reactions as defined in Section~\ref{sec:pdgf},
and (3) analyze the importance of individual reaction on downstream signaling molecules.
For the first goal, we study the signal transduction properties of each mutant
by removing the mutant arrows one by one and
examine how the states of each molecule change accordingly at different time instances.
We also examine the total time for
each molecule being active.
Moreover, it is interesting to study
the activities of each molecule at the steady state as well.
For the second goal, we do the comparison of probabilities
for molecules to be active between different mutants
by removing each of the crosstalk reactions.
For the last one, we study how the steady state probabilties of molecule {\tt MEK12} and {\tt Akt} change
when a certain reaction is removed.

Below we list properties of the PRISM model
that we have analyzed to achieve our goals.
Here, we use only the molecule {\tt PDGFR} to illustrate
the specification of the properties expressed as CSL formulas.
\begin{itemize}

\item $P_{=?}[\,F^{[t,t]} {\tt PDGFR}=1\,]$ \\
The probability that the molecule {\tt PDGFR} is active at time instant $t$.

\item $R^{\{``PDGFRactive"\}}_{=?}[C<=t]$ \\
The expected time of {\tt PDGFR} being active by time $t$.
It refers to the reward structure ``PDGFRactive" as defined in Figure~\ref{fig:rewards}.

\item $ S_{=?} [\,{\tt PDGFR}=1\,] $ \\
The long-run probability that {\tt PDGFR} is active.

\end{itemize}

\section{Results}
\label{sec:results}

\begin{figure}
  \subfigure[Probability of being active ({\tt PDGFR})\label{fig:mutantPDGFR}]{
            \begin{minipage}[c]{0.49\textwidth}
            \centering
            \includegraphics[width=1\textwidth]{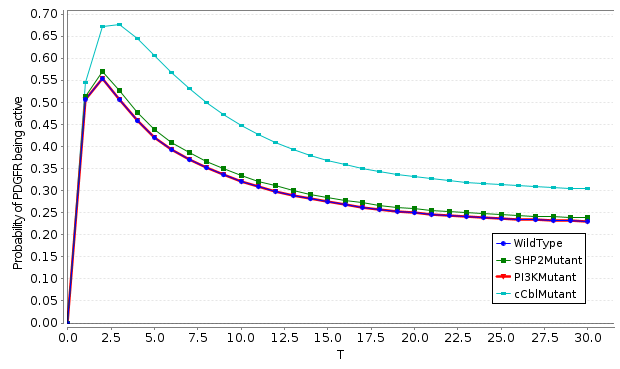}
            \end{minipage}}
\subfigure[Probability of being active ({\tt SHP2})\label{fig:mutantSHP2}]{
            \centering
            \begin{minipage}[c]{0.49\textwidth}
            \centering
            \includegraphics[width=1\textwidth]{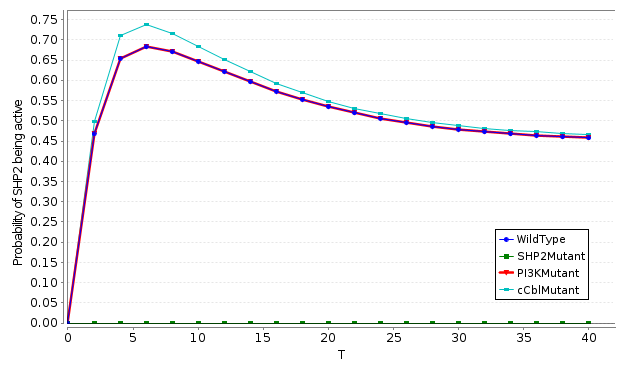}
            \end{minipage}}
\subfigure[Probability of being active ({\tt Ras})\label{fig:mutantRas}]{
            \centering
            \begin{minipage}[c]{0.49\textwidth}
            \centering
            \includegraphics[width=1\textwidth]{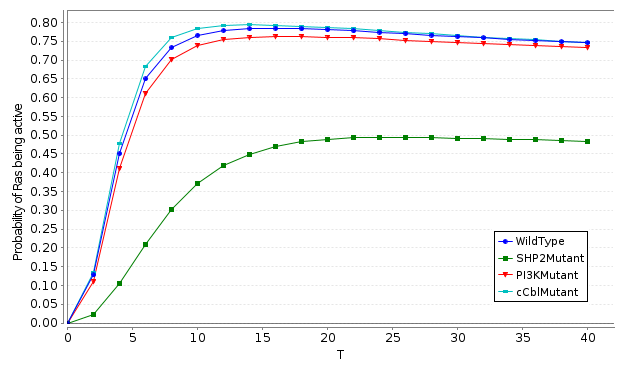}
            \end{minipage}}
\subfigure[Probability of being active ({\tt PIP3})\label{fig:mutantPIP3}]{
            \centering
            \begin{minipage}[c]{0.49\textwidth}
            \centering
            \includegraphics[width=1\textwidth]{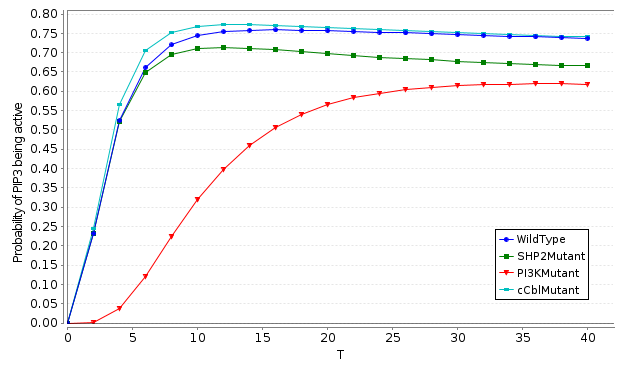}
            \end{minipage}}
\subfigure[Probability of being active ({\tt MEK12})\label{fig:mutantMEK12}]{
 	    \centering
            \begin{minipage}[c]{0.49\textwidth}
            \centering
            \includegraphics[width=1\textwidth]{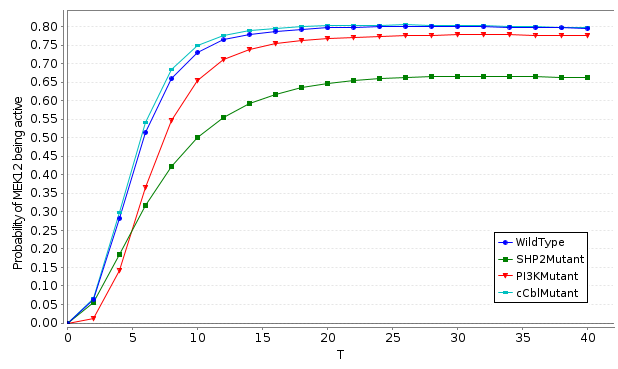}
            \end{minipage}}
\subfigure[Probability of being active ({\tt Akt})\label{fig:mutantAkt}]{
  	    \centering
            \begin{minipage}[c]{0.49\textwidth}
            \centering
            \includegraphics[width=1\textwidth]{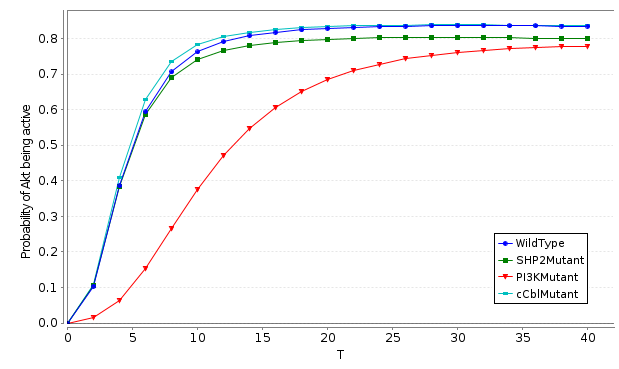}
            \end{minipage}}
\caption{Probabilities of molecules being active\label{fig:four-mutants} }
\end{figure}

We use PRISM to construct the PDGF model described in Section~\ref{ssec:modelling} and
analyze the set of properties listed in Section~\ref{ssec:properties}.
As described in Section~\ref{sec:pdgf},

For the first goal to analyze the dynamics of PDGF induced signaling, 
we first develop a base model representing the system,
in which all the reactions in Figure~\ref{fig:reactions} are included.
Subsequently, we obtain the mutant models by removing the mutant arrows
(as mentioned in Section~\ref{sec:pdgf}) one by one.
More precisely, we have developed four models in this experiment
including the base model corresponding to the {\tt WildType} condition.
The second one, called {\tt SHP2Mutant},
is obtained by removing the mutant arrow pointed to {\tt SHP2}.
The third and the fourth ones are {\tt PI3KMutant} and {\tt cCblMutant}.
Following the same process, we remove the mutant arrows pointed to {\tt PI3K} and {\tt cCbl} separately in the last two models.
The three removed mutant arrows correspond to the 
reactions (7), (8) and (6) in Figure~\ref{fig:reactions}.
The size of the models in PRISM are shown in Table~\ref{tab:sizestatistics}
(Table~\ref{tab:sizestatistics} also shows the size of models {\tt RasPI3KMutant} and {\tt Aktc-RafMutant}
which will be discussed later).
For each model, we compute the probability of each molecule being active at time instance $t$,
which is summarized in Figure~\ref{fig:four-mutants}.

\begin{table}
\begin{center}
  \begin{tabular}{ | l | r | r | }\hline
  	\hspace{10mm}\textbf{Model} \hspace{10mm} & \hspace{1.5cm} {\bf States} & \hspace{1cm} {\bf Transitions} \\ \hline\hline
	{\tt WildType} & 589,824 & 7,145,472 \\ \hline\hline
	{\tt SHP2Mutant} & 36,864 & 373,248 \\ \hline
	{\tt PI3KMutant} & 589,824 & 7,096,320 \\ \hline	
	{\tt cCb1Mutant} & 294,912 & 3,357,696 \\ \hline	\hline
	{\tt RasPI3KMutant} & 589,824 & 7,047,168 \\ \hline
	{\tt Aktc-RafMutant} & 786,423 & 10,264,576 \\ \hline					
  \end{tabular}
\end{center}
\caption{Model statistics in PRISM\label{tab:sizestatistics}}
\end{table}


Figure~\ref{fig:four-mutants} shows the probability of 6 out of the 17 molecules
namely {\tt PDGFR}, {\tt SHP2}, {\tt Ras}, {\tt PIP3}, {\tt MEK12}, and {\tt Akt}.
The 6 molecules are chosen according to their positions in the signaling pathway.
We can see from Figure~\ref{fig:mutantPDGFR} that {\tt cCbl} affects {\tt PDGFR}
more than the other two molecules ({\tt SHP2} and {\tt PI3K}).
This is due to strong negative feedback (red arrow) from {\tt cCbl} to {\tt PDGFR} in Figure~\ref{fig:pdgf}.
There is also a negative feedback from {\tt SHP2} to {\tt PDGFR}.
However, the reaction rate of the negative feedback from {\tt cCbl} is 7 times
as large as the one from {\tt SHP2}, hence {\tt cCbl} can affect {\tt PDGFR} more than {\tt SHP2}.
The red curve in Figure~\ref{fig:mutantPIP3} shows that
not only {\tt PIP3} in {\tt PI3KMutant} is less active than in other conditions,
but also the time of it becoming active is delayed.
In biological experiments, this delay is \emph{not} observed.
This is due to the fact that in the full PDGF signaling pathway model
there is another molecule {\tt PLCg} giving a fast positive crosstalk to {\tt PI3K},
which is removed in our simplified PDGF model here.

If we focus on the light blue curve of Figures~\ref{fig:mutantPDGFR} -~\ref{fig:mutantAkt},
we can see that {\tt cCbl} has much impact on {\tt PDGFR} but little impact on the other molecules.
This is because after {\tt PDGFR} activates {\tt SHP2} and {\tt PI3K},
the states of the other molecules are determined by {\tt SHP2} and {\tt PI3K}.
Furthermore, from the green curve of Figures~\ref{fig:mutantSHP2},~\ref{fig:mutantRas} and~\ref{fig:mutantMEK12},
we see that {\tt SHP2} has great effect on the status of the molecules on the MAPK pathway.
However, the more downstream the molecule is,
the less prominent the effect becomes (the effect to {\tt MEK12} is less than that to {\tt Ras}).
Due to the influence of the positive feedback from {\tt PI3K} and {\tt PIP3},
the molecules in the MAPK pathway can also become active without {\tt SHP2},
which is the result from positive crosstalk interactions.
Besides, the red curves of Figures~\ref{fig:mutantPIP3} and ~\ref{fig:mutantAkt} show
that the influence of {\tt PI3K} to the molecules in the PI3K/Akt  pathway,
which is similar to that of {\tt SHP2} to the MAPK pathway.

We analyze the long-run probability of the molecules being active after {\tt PDGFL} stimulation (shown in Table~\ref{tab:steady}).
The result demonstrates the activities of each molecule in this model at steady state.

\begin{table}[htbp]
\begin{center}
  \begin{tabular}{ | l | r || l | r | }\hline
  	{\bf Molecule} & {\bf Probability} & {\bf Molecule} & {\bf Probability} \\ \hline\hline
	{\tt PDGFR} & 0.22 & {\tt Grb2SOS} & 0.55\\ \hline
	{\tt SHP2} & 0.45 & {\tt Ras} & 0.72\\ \hline
	{\tt GabSOS} & 0.53 & {\tt MEK12} & 0.77\\ \hline
	{\tt c-Raf} & 0.63 & {\tt PIP3} & 0.72\\ \hline
	{\tt PI3K} & 0.62 & {\tt Akt} & 0.82\\ \hline
	{\tt PDK} & 0.83 & {\tt MTOR} & 0.84 \\ \hline
	{\tt cCbl} & 0.47 &  {\tt PPX} & 0.00\\ \hline
  \end{tabular}
\end{center}
\caption{Steady state probabilities of molecules\label{tab:steady}}
\end{table}

\begin{figure}
  \centering
  \includegraphics[width=0.7\textwidth]{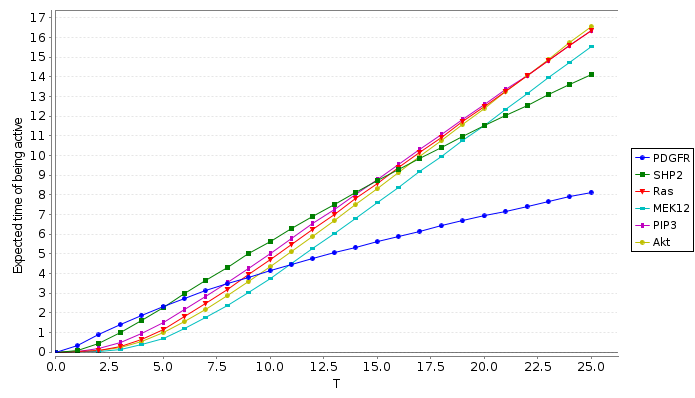}
  \caption{Expected time of being active by time $t$ \label{fig:reward}}
\end{figure}

PRISM supports reward properties (see Section~\ref{sec:modellingandproperty}).
Figure~\ref{fig:reward} shows the expected time of six molecules being active by time instant $t$.
These six molecules are the same as in Figure~\ref{fig:four-mutants}.
All the six curves tend to be linear after time instant 12,
which shows that the state of the 6 molecules start to be in a steady
state after time instant 12.

For the second goal to analyze the influence of the crosstalk reactions,
the experiment is also performed as in silico genetics.
After developing the base model, we get the model variants by removing one crosstalk arrow at a time.
Totally, there are four positive crosstalk reactions (green arrows) and two negative crosstalk reactions (red arrows pointing from Akt).
In this analysis, we focus on the positive crosstalk from {\tt Ras} to {\tt PI3K} and
the negative crosstalk from {\tt Akt} to {\tt c-Raf}.
More precisely, we develop three models in this experiment.
The first one is the base {\tt WildType} model with all reactions included.
The second one is the {\tt RasPI3KMutant} model,
in which the positive feedback from {\tt Ras} to {\tt PI3K} is removed.
The last one is the {\tt Aktc-RafMutant} model,
in which the negative feedback from {\tt Akt} to {\tt c-Raf} is removed.
After building the three models, we analyze the influences of the two crosstalks
by comparing the {\tt WildType} model to the two mutant models, respectively.
The size of each PRISM model is summarized in Table~\ref{tab:sizestatistics}.

\begin{figure}
  \subfigure[Influence of positive crosstalk from {\tt Ras} to {\tt PI3K}]{
    \label{fig:RasPI3KMutant} 
            \begin{minipage}[c]{0.49\textwidth}
               \centering
             \includegraphics[width=1\textwidth]{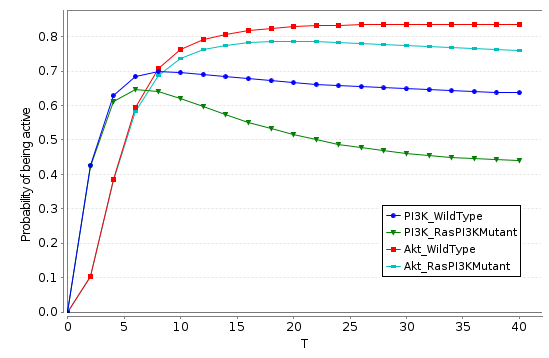}
            \end{minipage}}
\subfigure[Influence of negative crosstalk arrow from {\tt Akt} to {\tt c-Raf}]{
    \label{fig:Aktc-RafMutant} 
  	\centering
            \begin{minipage}[c]{0.49\textwidth}
            \centering
             \includegraphics[width=1\textwidth]{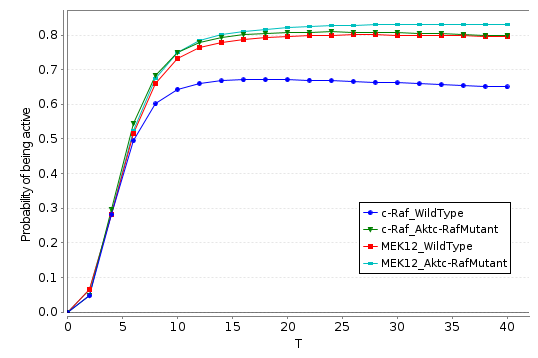}
            \end{minipage}}
  \caption{Influence of crosstalks \label{fig:results-crosstalks}}
\end{figure}

Figure~\ref{fig:results-crosstalks} shows the results of the comparison.
In both sub-figures, we compare the molecules
which are directly related to the crosstalk arrows and the molecules at the end of the related main signaling pathways.
In Figure~\ref{fig:RasPI3KMutant}, we compare in each model
the probabilities for molecules {\tt PI3K} and {\tt Akt} being active,
which is at the end of the PI3K/Akt pathway;
while in Figure~\ref{fig:Aktc-RafMutant} we have chosen the molecules {\tt c-Raf} and {\tt MEK12},
which is at the end of the MAPK pathway.
The dark blue and green curves in Figure~\ref{fig:RasPI3KMutant} show
how the green arrow influences the status of {\tt PI3K}.
If there is no positive crosstalk, the probability of {\tt PI3K} being active (green curve) becomes smaller.
The two curves are coincident for a time period
because the molecule {\tt Ras} needs some time to become active before it can activate {\tt PI3K}.
The red and light blue curves show that the influence of the
positive feedback to {\tt Akt} is smaller than that to {\tt PI3K}.
Just like the situation in Figures~\ref{fig:mutantSHP2},~\ref{fig:mutantRas} and ~\ref{fig:mutantMEK12}, the more downstream the molecules are,
the smaller the influence becomes.
The results in Figure~\ref{fig:Aktc-RafMutant} are similar:
the dark blue and green curves show
that the influence the negative crosstalk gives to {\tt Ras} is significant;
while the positive crosstalk has little influence on {\tt MEK12}.

\begin{figure}
  \centering
  \includegraphics[width=0.95\textwidth]{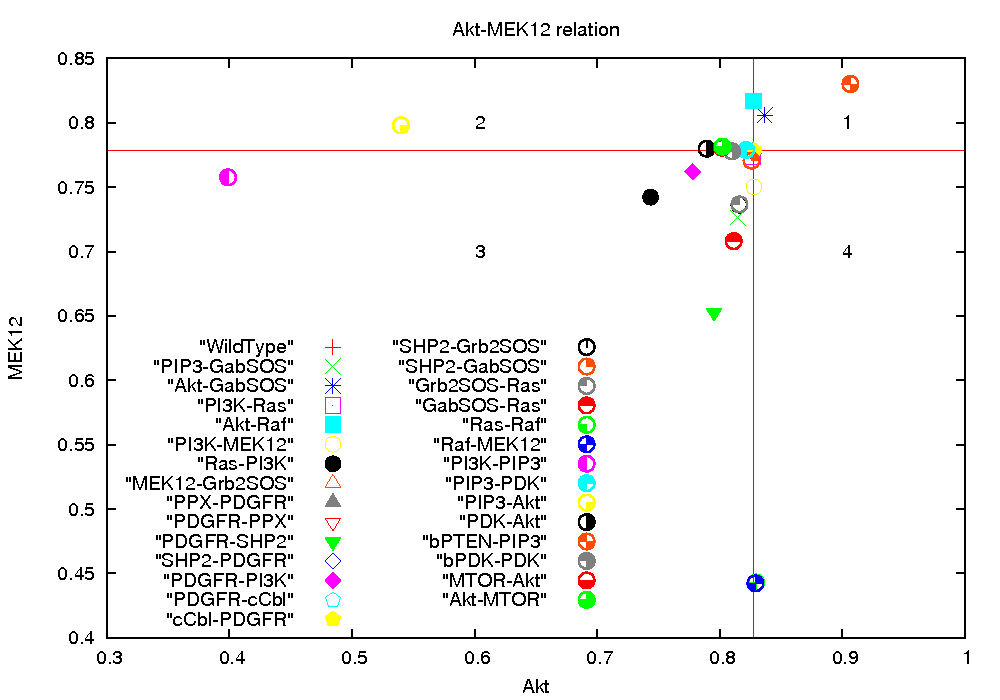}
  \caption{Steady state probabilities of {\tt Akt} and {\tt MEK12} in different models\label{fig:Akt-MEK12}}
\end{figure}

For the last goal to analyze the importance of individual reaction on downstream signaling molecules,
we compute the steady state probabilities of {\tt Akt} and {\tt MEK12} in 31 different models,
each of which is obtained by removing one reaction from the {\tt wildtype} model.
For example, `PIP3-GabSOS' model is obtained by removing reaction No.20 in Figure~\ref{fig:reactions}).
The results are shown in Figure~\ref{fig:Akt-MEK12}.
We draw two lines in the figure to divide the axis into four areas,
letting the dot of the wildtype model lie at the cross of the two lines.
Apparently, the dots in different areas show that the removed reactions can bring different influence to
the steady state probabilities of {\tt Akt} and {\tt MEK12}.
The reactions (corresponding to the dots) in area 1 can decrease the steady state probabilities of both {\tt Akt} and {\tt MEK12},
while those in area 3 can increase both probabilities.
The reactions in area 2 can decrease the steady state probability of {\tt Akt} and increase the one of {\tt MEK12};
while reactions grouped in area 4 lead to the opposite effect.
For those dots lying on the horizontal line,
the corresponding removed reactions have little impact on the steady state probability of {\tt Akt}.
Similarly, the reactions on the vertical line have little impact on {\tt MEK12}.
These observations have biological implications.
{\tt Akt} and {\tt MEK12} are downstream molecules in the signal transduction process
that regulate different cellular functions.
Signal from {\tt Akt} keeps the cells to survive from apoptosis and signal form {\tt MEK12} regulates the cells growth and proliferation.
In cancer, both of these two main pathways (see Figure~\ref{fig:pdgf})
are more active so they drive the cells to keep growing and dividing in an uncontrolled manner.
Therefore, if we could find the targets to control these two signaling molecules to be at a desirable level,
it would be beneficial for cancer therapeutic development.

\section{Conclusion and Future Work}
\label{sec:conclusion}

In this paper, we have given a detailed description of
using the probabilistic model checker PRISM to study the PDGF signaling pathway.
Based on intensive literature reviews, we have built a PDGF signal transduction model in ODE format
and extracted it for the analysis in a probabilistic model checker.
We focused on a simplified model of the PDGF signaling pathway which contains 17 molecules
while the full model has 35 molecules.
We have analyzed the  dynamics of PDGF induced signaling,
the influence of crosstalk reactions in the signaling pathway
and the importance of individual reaction on downstream signaling molecules.
Our experiment results show that
quantitative verification can provide us a better understanding of the PDGF signaling pathway,
especially the result discussed in the end of Sect.~\ref{sec:results}
potentially can give rise to better behavior prediction of the pathway.
(The PRISM model and property specifications can be found at \url{satoss.uni.lu/jun/models/PDGF.zip}.)

\begin{figure}
\subfigure[Probability of being active ({\tt PDGFR}) \label{fig:ODE_PDGFR}]{
 	    \centering
            \begin{minipage}[c]{0.49\textwidth}
            \centering
            \includegraphics[width=1\textwidth]{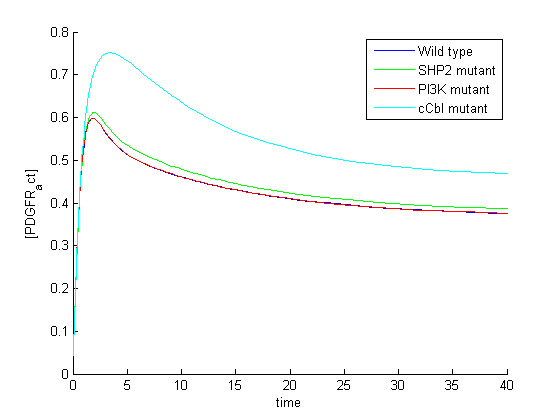}
            \end{minipage}}
\subfigure[Probability of being active ({\tt Ras}) \label{fig:ODE_Ras}]{
  	    \centering
            \begin{minipage}[c]{0.49\textwidth}
            \centering
            \includegraphics[width=1\textwidth]{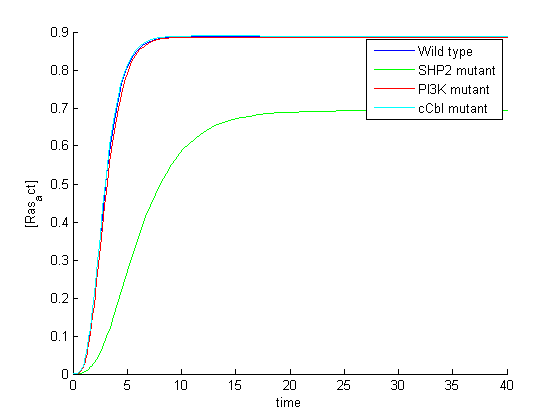}
            \end{minipage}}
\caption{Results from ODE simulations \label{fig:results-ode}}
\end{figure}
Except for model checking,
we have also analyzed the model using ODE simulations.
The results for the two molecules, {\tt PDGFR} and {\tt Ras},
are shown in Figure~\ref{fig:ODE_PDGFR} and Figure~\ref{fig:ODE_Ras} respectively.
We can see that probabilistic verification and ODE simulations produce comparable results:
the shape of the curves are similar while the maximum probabilities and the steady state probabilities differ a little.
The reason of the difference might be that the approach based on verification leads to more accurate results when
the number of molecules is small as indicated in~\cite{KNPTHG06}.
In general, ODE analysis perform well for larger number of molecules, but not for small numbers,
as time trajectories for average concentrations can be misleading if numbers of molecules small.
We plan to have a detailed comparison
between our analysis using probabilisitic verification techniques
and the analysis using ODE simulations.
As indicated in the end of Section~\ref{sec:pdgf},
our model is not intended to be fully accurate.
The reaction rates, especially the crosstalk reaction rates,
are based on literature reviewing and are only relative values.
Currently, experiments in a biological laboratory are performed
to get more precise values for these reaction rates.
We hope techniques like parametric verification for Markov chains (e.g.,~\cite{HKM08,HHZ09})
can help to synthesize values which are consistent with the lab experimental results.

\bibliographystyle{eptcs}

\begin{thebibliography}{10}
\providecommand{\bibitemdeclare}[2]{}
\providecommand{\urlprefix}{Available at }
\providecommand{\url}[1]{\texttt{#1}}
\providecommand{\href}[2]{\texttt{#2}}
\providecommand{\urlalt}[2]{\href{#1}{#2}}
\providecommand{\doi}[1]{doi:\urlalt{http://dx.doi.org/#1}{#1}}
\providecommand{\bibinfo}[2]{#2}

\bibitemdeclare{article}{KTM+96}
\bibitem{KTM+96}
\bibinfo{author}{K.~Anan}, \bibinfo{author}{T.~Morisaki},
  \bibinfo{author}{M.~Katano}, \bibinfo{author}{A.~Ikubo},
  \bibinfo{author}{H.~Kitsuki}, \bibinfo{author}{A.~Uchiyama},
  \bibinfo{author}{S.~Kuroki}, \bibinfo{author}{M.~Tanaka} \&
  \bibinfo{author}{M.~Torisu} (\bibinfo{year}{1996}):
  \emph{\bibinfo{title}{Vascular endothelial growth factor and platelet-derived
  growth factor are potential angiogenic and metastatic factors in human breast
  cancer}}.
\newblock {\sl \bibinfo{journal}{Surgery}}
  \bibinfo{volume}{119}(\bibinfo{number}{3}), pp. \bibinfo{pages}{333--339}.

\bibitemdeclare{article}{ASSB00}
\bibitem{ASSB00}
\bibinfo{author}{A.~Aziz}, \bibinfo{author}{K.~Sanwal},
  \bibinfo{author}{V.~Singhal} \& \bibinfo{author}{R.~Brayton}
  (\bibinfo{year}{2000}): \emph{\bibinfo{title}{Model Checking Continuous Time
  Markov Chains}}.
\newblock {\sl \bibinfo{journal}{ACM Transactions on Computational Logic}}
  \bibinfo{volume}{1}(\bibinfo{number}{1}), pp. \bibinfo{pages}{162--170}.
    \doi{10.1145/343369.343402}

\bibitemdeclare{article}{BGK96}
\bibitem{BGK96}
\bibinfo{author}{C.~E. Bazenet}, \bibinfo{author}{J.~A. Gelderloos} \&
  \bibinfo{author}{A.~Kazlauskas} (\bibinfo{year}{1996}):
  \emph{\bibinfo{title}{{Phosphorylation of tyrosine 720 in the
  platelet-derived growth factor alpha receptor is required for binding of Grb2
  and SHP-2 but not for activation of Ras or cell proliferation}}}.
\newblock {\sl \bibinfo{journal}{Molecular and Cellular Biology}}
  \bibinfo{volume}{16}(\bibinfo{number}{12}), pp. \bibinfo{pages}{6926--6936}.

\bibitemdeclare{article}{Bha03}
\bibitem{Bha03}
\bibinfo{author}{U.S. Bhalla} (\bibinfo{year}{2003}):
  \emph{\bibinfo{title}{Understanding complex signaling networks through models
  and metaphor}}.
\newblock {\sl \bibinfo{journal}{Progress in Biophysics \& Molecular Biology}}
  \bibinfo{volume}{81}(\bibinfo{number}{1}), pp. \bibinfo{pages}{45--65}.

\bibitemdeclare{inproceedings}{BFF+09}
\bibitem{BFF+09}
\bibinfo{author}{N.~Bonzanni}, \bibinfo{author}{K.~{Anton Feenstra}},
  \bibinfo{author}{W.~J. Fokkink} \& \bibinfo{author}{E.~Krepska}
  (\bibinfo{year}{2009}): \emph{\bibinfo{title}{What Can Formal Methods Bring
  to Systems Biology?}}
\newblock In: {\sl \bibinfo{booktitle}{Proc.\ 2nd World Congress on Formal
  Methods}}, {\sl \bibinfo{series}{LNCS}} \bibinfo{volume}{5850},
  \bibinfo{publisher}{Springer}, pp. \bibinfo{pages}{16--22}.
  \doi{10.1007/978-3-642-05089-3\_2}

\bibitemdeclare{article}{BKF+09}
\bibitem{BKF+09}
\bibinfo{author}{N.~Bonzanni}, \bibinfo{author}{E.~Krepska},
  \bibinfo{author}{K.~{Anton Feenstra}}, \bibinfo{author}{W.~J. Fokkink},
  \bibinfo{author}{T.~Kielmann}, \bibinfo{author}{H.~E. Bal} \&
  \bibinfo{author}{J.~Heringa} (\bibinfo{year}{2009}):
  \emph{\bibinfo{title}{Executing multicellular differentiation: quantitative
  predictive modelling of {\it C.elegans} vulval development}}.
\newblock {\sl \bibinfo{journal}{Bioinformatics}}
  \bibinfo{volume}{25}(\bibinfo{number}{16}), pp. \bibinfo{pages}{2049--2056}.
  \doi{10.1093/bioinformatics/btp355}

\bibitemdeclare{article}{CDHH10}
\bibitem{CDHH10}
\bibinfo{author}{F.~Ciocchetta}, \bibinfo{author}{A.~Degasperi},
  \bibinfo{author}{J.~K. Heath} \& \bibinfo{author}{J.~Hillston}
  (\bibinfo{year}{2010}): \emph{\bibinfo{title}{Modelling and Analysis of the
  NF-{\it appa}B Pathway in {B}io-{PEPA}}}.
\newblock {\sl \bibinfo{journal}{Transactions on Computational Systems
  Biology}} \bibinfo{volume}{12}, pp. \bibinfo{pages}{229--262}.
  \doi{10.1007/978-3-642-11712-1\_7}

\bibitemdeclare{article}{JDS+03}
\bibitem{JDS+03}
\bibinfo{author}{J.~Cools}, \bibinfo{author}{D.~J. DeAngelo},
  \bibinfo{author}{J.~Gotlib}, \bibinfo{author}{E.~H. Stover},
  \bibinfo{author}{R.~D. Legare}, \bibinfo{author}{J.~Cortes},
  \bibinfo{author}{J.~Kutok}, \bibinfo{author}{J.~Clark},
  \bibinfo{author}{I.~Galinsky}, \bibinfo{author}{J.~D. Griffin},
  \bibinfo{author}{N.~C. Cross}, \bibinfo{author}{A.~Tefferi},
  \bibinfo{author}{J.~Malone}, \bibinfo{author}{R.~Alam},
  \bibinfo{author}{S.~L. Schrier}, \bibinfo{author}{J.~Schmid},
  \bibinfo{author}{M.~Rose}, \bibinfo{author}{P.~Vandenberghe},
  \bibinfo{author}{G.~Verhoef}, \bibinfo{author}{M.~Boogaerts},
  \bibinfo{author}{I.~Wlodarska}, \bibinfo{author}{H.~Kantarjian},
  \bibinfo{author}{P.~Marynen}, \bibinfo{author}{S.~E. Coutre},
  \bibinfo{author}{R.~Stone} \& \bibinfo{author}{D.~G. Gilliland}
  (\bibinfo{year}{2003}): \emph{\bibinfo{title}{{A tyrosine kinase created by
  fusion of the PDGFRA and FIP1L1 genes as a therapeutic target of imatinib in
  idiopathic hypereosinophilic syndrome}}}.
\newblock {\sl \bibinfo{journal}{The New England journal of medicine}}
  \bibinfo{volume}{348}(\bibinfo{number}{13}), pp. \bibinfo{pages}{1201--1214}.

\bibitemdeclare{inproceedings}{DMM10}
\bibitem{DMM10}
\bibinfo{author}{P.~Dr{\'a}bik}, \bibinfo{author}{A.~Maggiolo-Schettini} \&
  \bibinfo{author}{P.~Milazzo} (\bibinfo{year}{2010}):
  \emph{\bibinfo{title}{Modular verification of interactive systems with an
  application to biology}}.
\newblock In: {\sl \bibinfo{booktitle}{Proc.\ 1st Workshop on Interactions
  between Computer Science and Biology}}, {\sl \bibinfo{series}{ENTCS}}
  \bibinfo{volume}{268}, \bibinfo{publisher}{Elsevier}, pp.
  \bibinfo{pages}{61--75}.
  \doi{10.1016/j.entcs.2010.12.006}

\bibitemdeclare{article}{FH07}
\bibitem{FH07}
\bibinfo{author}{J.~Fisher} \& \bibinfo{author}{T.~A. Henzinger}
  (\bibinfo{year}{2007}): \emph{\bibinfo{title}{Executable cell biology}}.
\newblock {\sl \bibinfo{journal}{Nature Biotechnology}}
  \bibinfo{volume}{25}(\bibinfo{number}{11}), pp. \bibinfo{pages}{1239--1249}.

\bibitemdeclare{article}{FP10}
\bibitem{FP10}
\bibinfo{author}{J.~Fisher} \& \bibinfo{author}{N.~Piterman}
  (\bibinfo{year}{2010}): \emph{\bibinfo{title}{The Executable Pathway to
  Biological Networks}}.
\newblock {\sl \bibinfo{journal}{Briefings in Functional Genomics and
  Proteomics}} \bibinfo{volume}{9}(\bibinfo{number}{1}), pp.
  \bibinfo{pages}{79--92}.

\bibitemdeclare{article}{GBLG94}
\bibitem{GBLG94}
\bibinfo{author}{T.~R. Golub}, \bibinfo{author}{G.~F. Barker},
  \bibinfo{author}{M.~Lovett} \& \bibinfo{author}{D.~G. Gilliland}
  (\bibinfo{year}{1994}): \emph{\bibinfo{title}{{Fusion of PDGF receptor beta
  to a novel ets-like gene, tel, in chronic myelomonocytic leukemia with
  t(5;12) chromosomal translocation}}}.
\newblock {\sl \bibinfo{journal}{Cell}}
  \bibinfo{volume}{77}(\bibinfo{number}{2}), pp. \bibinfo{pages}{307--316}.

\bibitemdeclare{article}{GAP10}
\bibitem{GAP10}
\bibinfo{author}{A.~Guti\'errez-Uzquiza}, \bibinfo{author}{M.~Arechederra},
  \bibinfo{author}{I.~Molina}, \bibinfo{author}{R.~Banos},
  \bibinfo{author}{V.~Maia}, \bibinfo{author}{M.~Benito},
  \bibinfo{author}{C.~Guerrero} \& \bibinfo{author}{A.~Porras}
  (\bibinfo{year}{2010}): \emph{\bibinfo{title}{{C3G down-regulates p38 MAPK
  activity in response to stress by Rap-1 independent mechanisms: involvement
  in cell death}}}.
\newblock {\sl \bibinfo{journal}{Cellualar Signalling}}
  \bibinfo{volume}{22}(\bibinfo{number}{3}), pp. \bibinfo{pages}{533--542}.

\bibitemdeclare{inproceedings}{HHZ09}
\bibitem{HHZ09}
\bibinfo{author}{E.~M. Hahn}, \bibinfo{author}{H.~Hermanns} \&
  \bibinfo{author}{L.~Zhang} (\bibinfo{year}{2009}):
  \emph{\bibinfo{title}{Probabilistic reachability for parametric {M}arkov
  models}}.
\newblock In: {\sl \bibinfo{booktitle}{Proc.\ 16th Spin Workshop on Model
  Checking Software}}, {\sl \bibinfo{series}{LNCS}} \bibinfo{volume}{5578},
  \bibinfo{publisher}{Springer}, pp. \bibinfo{pages}{88--106}.
  \doi{10.1007/978-3-642-02652-2\_10}

\bibitemdeclare{inproceedings}{HKM08}
\bibitem{HKM08}
\bibinfo{author}{T.~Han}, \bibinfo{author}{J.-P. Katoen} \&
  \bibinfo{author}{A.~Mereacre} (\bibinfo{year}{2008}):
  \emph{\bibinfo{title}{Approximate parameter synthesis for probabilistic
  time-bounded reachability}}.
\newblock In: {\sl \bibinfo{booktitle}{Proc.\ 29 IEEE Real-Time Systems
  Symposium}}, \bibinfo{publisher}{IEEE Computer Society}, pp.
  \bibinfo{pages}{173--182}.
  \doi{10.1109/RTSS.2008.19}

\bibitemdeclare{article}{HW11}
\bibitem{HW11}
\bibinfo{author}{D.~Hanahan} \& \bibinfo{author}{R.~A. Weinberg}
  (\bibinfo{year}{2011}): \emph{\bibinfo{title}{Hallmarks of cancer: The next
  generation}}.
\newblock {\sl \bibinfo{journal}{Cell}}
  \bibinfo{volume}{144}(\bibinfo{number}{5}), pp. \bibinfo{pages}{646--674}.

\bibitemdeclare{article}{HSP83}
\bibitem{HSP83}
\bibinfo{author}{S.~Hart}, \bibinfo{author}{M.~Sharir} \&
  \bibinfo{author}{A.~Pnueli} (\bibinfo{year}{1983}):
  \emph{\bibinfo{title}{Termination of probabilistic concurrent programs}}.
\newblock {\sl \bibinfo{journal}{ACM Transactions on Programming Languages and
  Systems}} \bibinfo{volume}{5}(\bibinfo{number}{3}), pp.
  \bibinfo{pages}{356--380}.
  \doi{10.1145/2166.357214}

\bibitemdeclare{article}{HKN+08}
\bibitem{HKN+08}
\bibinfo{author}{J.~Heath}, \bibinfo{author}{M.~Kwiatkowska},
  \bibinfo{author}{G.~Norman}, \bibinfo{author}{D.~Parker} \&
  \bibinfo{author}{O.~Tymchyshyn} (\bibinfo{year}{2008}):
  \emph{\bibinfo{title}{Probabilistic model checking of complex biological
  pathways}}.
\newblock {\sl \bibinfo{journal}{Theoretical Computer Science}}
  \bibinfo{volume}{319}(\bibinfo{number}{3}), pp. \bibinfo{pages}{239--257}.
  \doi{10.1016/j.tcs.2007.11.013}

\bibitemdeclare{article}{HWW85}
\bibitem{HWW85}
\bibinfo{author}{C.-H. Heldin}, \bibinfo{author}{B.~Westermarkt} \&
  \bibinfo{author}{A.~Wasteson} (\bibinfo{year}{1985}):
  \emph{\bibinfo{title}{Platelet-derived growth factor}}.
\newblock {\sl \bibinfo{journal}{Molecular and Celluar Endocrinology}}
  \bibinfo{volume}{39}(\bibinfo{number}{3}), pp. \bibinfo{pages}{169--187}.

\bibitemdeclare{inproceedings}{KCLLPZ09}
\bibitem{KCLLPZ09}
\bibinfo{author}{S.~K. Jha}, \bibinfo{author}{E.~M. Clarke},
  \bibinfo{author}{C.~J. Langmead}, \bibinfo{author}{A.~Legay},
  \bibinfo{author}{A.~Platzer} \& \bibinfo{author}{P.~Zuliani}
  (\bibinfo{year}{2009}): \emph{\bibinfo{title}{A Bayesian Approach to Model
  Checking Biological Systems}}.
\newblock In: {\sl \bibinfo{booktitle}{Proc.\ 7th Conference on Computational
  Methods in Systems Biology}}, {\sl \bibinfo{series}{LNCS}}
  \bibinfo{volume}{5688}, \bibinfo{publisher}{Springer}, pp.
  \bibinfo{pages}{218--234}.
  \doi{10.1007/978-3-642-03845-7\_15}

\bibitemdeclare{article}{JWL99}
\bibitem{JWL99}
\bibinfo{author}{C.~A. Joazeiro}, \bibinfo{author}{S.~S. Wing},
  \bibinfo{author}{H.~Huang}, \bibinfo{author}{J.~D. Leverson},
  \bibinfo{author}{T.~Hunter} \& \bibinfo{author}{Y.~C. Liu}
  (\bibinfo{year}{1999}): \emph{\bibinfo{title}{The tyrosine kinase negative
  regulator c-{C}bl as a {RING}-type, {E}2-dependent ubiquitin-protein
  ligase}}.
\newblock {\sl \bibinfo{journal}{Science}}
  \bibinfo{volume}{286}(\bibinfo{number}{5438}), pp. \bibinfo{pages}{309--312}.

\bibitemdeclare{book}{Kra08}
\bibitem{Kra08}
\bibinfo{author}{G.~Krauss} (\bibinfo{year}{2008}):
  \emph{\bibinfo{title}{Biochemistry of Signal Transduction and Regulation}}.
\newblock \bibinfo{publisher}{Wiley-VCH, Weinheim}.

\bibitemdeclare{inproceedings}{KDB09}
\bibitem{KDB09}
\bibinfo{author}{J.~Krivine}, \bibinfo{author}{V.~Danos} \&
  \bibinfo{author}{JA. Benecke} (\bibinfo{year}{2009}):
  \emph{\bibinfo{title}{Modelling epigenetic information maintenance: a Kappa
  tutorial}}.
\newblock In: {\sl \bibinfo{booktitle}{Proc.\ 21st Conference on Computer Aided
  Verification}}, {\sl \bibinfo{series}{LNCS}} \bibinfo{volume}{5643},
  \bibinfo{publisher}{Springer}, pp. \bibinfo{pages}{17--32}.
  \doi{10.1007/978-3-642-02658-4\_3}

\bibitemdeclare{article}{KNP08}
\bibitem{KNP08}
\bibinfo{author}{M.~Z. Kwiatkowska}, \bibinfo{author}{G.~Norman} \&
  \bibinfo{author}{D.~Parker} (\bibinfo{year}{2008}):
  \emph{\bibinfo{title}{Using probabilistic model checking in systems
  biology}}.
\newblock {\sl \bibinfo{journal}{{SIGMETRICS} Performance Evaluation Review}}
  \bibinfo{volume}{35}, pp. \bibinfo{pages}{14--21}.
  \doi{10.1145/1364644.1364651}

\bibitemdeclare{inbook}{KNP09b}
\bibitem{KNP09b}
\bibinfo{author}{M.~Z. Kwiatkowska}, \bibinfo{author}{G.~Norman} \&
  \bibinfo{author}{D.~Parker} (\bibinfo{year}{2009}):
  \emph{\bibinfo{title}{Algorithmic Bioprocesses}}, chapter
  \bibinfo{chapter}{Quantitative Verification Techniques for Biological
  Processes}, pp. \bibinfo{pages}{391--409}.
\newblock \bibinfo{publisher}{Springer}.

\bibitemdeclare{inbook}{KNP10a}
\bibitem{KNP10a}
\bibinfo{author}{M.~Z. Kwiatkowska}, \bibinfo{author}{G.~Norman} \&
  \bibinfo{author}{D.~Parker} (\bibinfo{year}{2010}):
  \emph{\bibinfo{title}{Symbolic Systems Biology}}, chapter
  \bibinfo{chapter}{Probabilistic Model Checking for Systems Biology}, pp.
  \bibinfo{pages}{31--59}.
\newblock \bibinfo{publisher}{Jones and Bartlett}.

\bibitemdeclare{inproceedings}{KNPTHG06}
\bibitem{KNPTHG06}
\bibinfo{author}{M.~Z. Kwiatkowska}, \bibinfo{author}{G.~Norman},
  \bibinfo{author}{D.~Parker}, \bibinfo{author}{O.~Tymchyshyn},
  \bibinfo{author}{J.~Heath} \& \bibinfo{author}{E.~Gaffney}
  (\bibinfo{year}{2006}): \emph{\bibinfo{title}{Simulation and verification for
  computational modelling of signalling pathways}}.
\newblock In: {\sl \bibinfo{booktitle}{Proc.\ 38th Winter Simulation
  Conference}}, pp. \bibinfo{pages}{1666--1674}.
  \doi{10.1145/1218112.1218415}

\bibitemdeclare{article}{ML06}
\bibitem{ML06}
\bibinfo{author}{M.~Miettinen} \& \bibinfo{author}{J.~Lasota}
  (\bibinfo{year}{2006}): \emph{\bibinfo{title}{Gastrointestinal stromal
  tumors: {P}athology and prognosis at different sites}}.
\newblock {\sl \bibinfo{journal}{Seminars in Diagnostic Pathology}}
  \bibinfo{volume}{23}(\bibinfo{number}{2}), pp. \bibinfo{pages}{70--83}.

\bibitemdeclare{article}{RYD+07}
\bibitem{RYD+07}
\bibinfo{author}{A.~L. Reddi}, \bibinfo{author}{G.~Ying},
  \bibinfo{author}{L.~Duan}, \bibinfo{author}{G.~Chen},
  \bibinfo{author}{M.~Dimri}, \bibinfo{author}{P.~Douillard},
  \bibinfo{author}{B.~J. Druker}, \bibinfo{author}{M.~Naramura},
  \bibinfo{author}{V.~Band} \& \bibinfo{author}{H.~Band}
  (\bibinfo{year}{2007}): \emph{\bibinfo{title}{{Binding of Cbl to a
  phospholipase Cgamma1-docking site on platelet-derived growth factor receptor
  beta provides a dual mechanism of negative regulation}}}.
\newblock {\sl \bibinfo{journal}{Journal of Biological Chemistry}}
  \bibinfo{volume}{282}(\bibinfo{number}{40}), pp.
  \bibinfo{pages}{29336--2947}.

\bibitemdeclare{article}{RS02}
\bibitem{RS02}
\bibinfo{author}{A.~Regev} \& \bibinfo{author}{E.~Shapiro}
  (\bibinfo{year}{2002}): \emph{\bibinfo{title}{Cellular abstractions: {C}ells
  as computation}}.
\newblock {\sl \bibinfo{journal}{Nature}} \bibinfo{volume}{419}, p.
  \bibinfo{pages}{343}.

\bibitemdeclare{article}{SFB+08}
\bibitem{SFB+08}
\bibinfo{author}{A.~Sadot}, \bibinfo{author}{J.~Fisher},
  \bibinfo{author}{D.~Barak}, \bibinfo{author}{Y.~Admanit},
  \bibinfo{author}{M.~J. Stern}, \bibinfo{author}{E.~{Jane Albert Hubbard}} \&
  \bibinfo{author}{D.~Harel} (\bibinfo{year}{2008}):
  \emph{\bibinfo{title}{Toward Verified Biological Models}}.
\newblock {\sl \bibinfo{journal}{IEEE/ACM Transactions on Computational Biology
  Bioinformatics}} \bibinfo{volume}{5}(\bibinfo{number}{2}), pp.
  \bibinfo{pages}{223--234}.
  \doi{10.1145/1371585.1371591}

\bibitemdeclare{inproceedings}{SH09}
\bibitem{SH09}
\bibinfo{author}{M.~Schwarick} \& \bibinfo{author}{M.~Heiner}
  (\bibinfo{year}{2009}): \emph{\bibinfo{title}{{CSL} Model Checking of
  Biochemical Networks with Interval Decision Diagram}}.
\newblock In: {\sl \bibinfo{booktitle}{Proc.\ 7th Conference on Computational
  Methods in Systems Biology}}, {\sl \bibinfo{series}{LNCS}}
  \bibinfo{volume}{5688}, \bibinfo{publisher}{Springer}, pp.
  \bibinfo{pages}{296--312}.
  \doi{10.1007/978-3-642-03845-7\_20}

\bibitemdeclare{article}{TK04}
\bibitem{TK04}
\bibinfo{author}{M.~Tallquist} \& \bibinfo{author}{A.~Kazlauskas}
  (\bibinfo{year}{2004}): \emph{\bibinfo{title}{{PDGF} signaling in cells and
  mice}}.
\newblock {\sl \bibinfo{journal}{Cytokine \& Growth Factor Review}}
  \bibinfo{volume}{15}(\bibinfo{number}{4}), pp. \bibinfo{pages}{205--213}.

\bibitemdeclare{article}{VPC98}
\bibitem{VPC98}
\bibinfo{author}{S.~Valgeirsd\'ottir}, \bibinfo{author}{K.~Paukku},
  \bibinfo{author}{O.~Silvennoinen}, \bibinfo{author}{C.~H. Heldin} \&
  \bibinfo{author}{L.~Claesson-Welsh} (\bibinfo{year}{1998}):
  \emph{\bibinfo{title}{{Activation of Stat5 by platelet-derived growth factor
  (PDGF) is dependent on phosphorylation sites in PDGF beta-receptor
  juxtamembrane and kinase insert domains}}}.
\newblock {\sl \bibinfo{journal}{Oncogene}}
  \bibinfo{volume}{16}(\bibinfo{number}{4}), pp. \bibinfo{pages}{505--515}.

\bibitemdeclare{article}{YHM98}
\bibitem{YHM98}
\bibinfo{author}{K.~Yokote}, \bibinfo{author}{U.~Hellman},
  \bibinfo{author}{S.~Ekman}, \bibinfo{author}{Y.~Saito},
  \bibinfo{author}{L.~Roennstrand}, \bibinfo{author}{Y.~Saito},
  \bibinfo{author}{C.~H. Heldin} \& \bibinfo{author}{S.~Mori}
  (\bibinfo{year}{1998}): \emph{\bibinfo{title}{{Identification of Tyr-762 in
  the platelet-derived growth factor alpha-receptor as the binding site for Crk
  proteins}}}.
\newblock {\sl \bibinfo{journal}{Oncogene}}
  \bibinfo{volume}{16}(\bibinfo{number}{10}), pp. \bibinfo{pages}{1229--1239}.

\bibitemdeclare{article}{YGM+94}
\bibitem{YGM+94}
\bibinfo{author}{J.~Yu}, \bibinfo{author}{J.~S. Gutkind},
  \bibinfo{author}{D.~Mahadevan}, \bibinfo{author}{W.~Li},
  \bibinfo{author}{K.~A. Meyers}, \bibinfo{author}{J.~H. Pierce} \&
  \bibinfo{author}{M.~A. Heidaran} (\bibinfo{year}{1994}):
  \emph{\bibinfo{title}{{Biological function of PDGF-induced PI-3 kinase
  activity: its role in alpha PDGF receptor-mediated mitogenic signaling}}}.
\newblock {\sl \bibinfo{journal}{Journal of Cell Biology}}
  \bibinfo{volume}{127}(\bibinfo{number}{2}), pp. \bibinfo{pages}{479--487}.

\bibitemdeclare{article}{YUK03}
\bibitem{YUK03}
\bibinfo{author}{J.~Yu}, \bibinfo{author}{C.~Ustach} \& \bibinfo{author}{H.-R.
  Kim} (\bibinfo{year}{2003}): \emph{\bibinfo{title}{Platelet-derived growth
  factor signaling and human cancer}}.
\newblock {\sl \bibinfo{journal}{Journal of Biochemistry and Molecular
  Biology}} \bibinfo{volume}{36}(\bibinfo{number}{1}), pp.
  \bibinfo{pages}{49--59}.

\end{thebibliography}

\end{document}